\documentclass[twocolumn,preprintnumbers,prd,floatfix,nofootinbib]{revtex4}
\pdfoutput=1
\usepackage{hyperref}
\usepackage{microtype}
\usepackage{footmisc,multirow}
\usepackage{subfigure}
\usepackage{amsmath,slashed}
\usepackage{amssymb}
\usepackage{graphicx}
\usepackage{multirow}
\usepackage{xcolor}
\usepackage{xspace}
\usepackage{booktabs}

\AtBeginDocument{
\heavyrulewidth=.08em
\lightrulewidth=.05em
\cmidrulewidth=.03em
\belowrulesep=.65ex
\belowbottomsep=0pt
\aboverulesep=.4ex
\abovetopsep=0pt
\cmidrulesep=\doublerulesep
\cmidrulekern=.5em
\defaultaddspace=.5em
}

\newcommand{\be}{\begin{eqnarray*}}
\newcommand{\ee}{\end{eqnarray*}}

\newcommand{\bee}{\begin{eqnarray}}
\newcommand{\eee}{\end{eqnarray}}
\newcommand{\beeq}{\begin{equation}}
\newcommand{\eeeq}{\end{equation}}

\newcommand{\tev}{\ensuremath{\text{T}\mspace{0.2mu}\text{e}\mspace{-1mu}\text{V}}\xspace}

\begin{document}

\title{A global fit of top quark effective theory to data}

\begin{abstract}
  In this paper we present a global fit of beyond the Standard Model
  (BSM) dimension six operators relevant to the top quark sector to all
  currently available top production cross-section measurements, namely
  parton-level top-pair and single top production at the LHC and the
  Tevatron. Higher order QCD corrections are modelled using
  differential and global $K$-factors, and we use novel fast-fitting
  techniques developed in the context of Monte Carlo event generator
  tuning to perform the fit. This allows us to provide new, fully
  correlated and model-independent bounds on new physics effects in
  the top sector from the most current direct hadron-collider
  measurements in light of the involved theoretical and experimental
  systematics. As a by-product, our analysis constitutes a
  proof-of-principle that fast fitting of theory to data is possible
  in the top quark sector, and paves the way for a more detailed
  analysis including top quark decays, detector corrections and
  precision observables. 
\end{abstract}

\author{Andy~Buckley}
\author{Christoph~Englert}
\author{James~Ferrando}
\author{David~J.~Miller}
\author{Liam~Moore}
\author{Michael~Russell}
\author{Chris~D.~White}
\collaboration{\textsc{TopFitter} Collaboration}
\affiliation{SUPA, School of Physics and Astronomy, University of
  Glasgow,\\Glasgow, G12 8QQ, UK}

\preprint{GLAS-PPE/2015-04}

\maketitle

\section{Introduction}
\label{sec:intro}
The Standard Model of particle physics has proven to be an extremely
successful description of Nature up to the electroweak
scale. Nonetheless there are many compelling reasons to believe it is
an intermediate step to a more fundamental picture of physics at the
\tev scale.

The top quark, as the heaviest Standard Model particle, is expected to
play a unique role in this new physics. Given the unsatisfactory
explanation of electroweak symmetry breaking within the SM and the
appearance of $m_t$ at the electroweak scale, i.e. the closeness of
the top Yukawa coupling to unity, the top mass may arguably be seen as a strong
hint of physics beyond the SM.

Most BSM scenarios lend a special role to the top quark. In
supersymmetry the light Higgs mass is stabilised from UV divergences
by the contribution of SUSY top partners, among others (see, e.g.,
Refs. \cite{Nilles:1983ge,Haber:1984rc}). In compositeness
scenarios~\cite{Contino:2003ve,Agashe:2004rs}, the quark masses and
EWSB are generated through linear couplings of the SM fermions to new
strongly-interacting physics at the \tev scale. In theories of warped
extra-dimensions, the top quark couples preferentially to Kaluza-Klein
states in the 5D bulk~\cite{ArkaniHamed:1999dc,Gherghetta:2000qt},
offering a unique window to the new physics.

Typically all these scenarios predict a modification of Higgs
phenomenology, which has been thoroughly studied after the Higgs
discovery~\cite{Corbett:2015ksa,Ellis:2014jta,Contino:2013kra,Pomarol:2013zra,Efrati:2015eaa}. Such
analyses are currently limited by small statistics in the observed
Higgs discovery modes. Taking the special role of the top quark in
electroweak symmetry breaking at face value, the abundant production
of top quarks at the LHC provides a complementary avenue to search for
new non-resonant physics beyond the SM, that will be relevant to our
understanding of electroweak symmetry breaking.

\begin{table}[!b]
\begin{tabular}{ll@{\qquad}ll}
\toprule
\multicolumn{2}{l}{4-fermion operators} & \multicolumn{2}{l}{Non 4-fermion operators} \\
\midrule
$O^{1}_{qq}$ & ($\bar{q}\gamma_{\mu}q)( \bar{q}\gamma^{\mu}q)$& $O^{3}_{\phi q}$  & $i(\phi^\dagger \tau^I D_\mu \phi )(\bar{q}\gamma^\mu \tau^I q) $ \\  
$O^{3}_{qq}$ &  ($\bar{q}\gamma_{\mu}\tau^Iq)( \bar{q}\gamma^{\mu}\tau^I q)$ & $O_{tW}$ & $ (\bar{q}\sigma^{\mu \nu} \tau^I t)\tilde \phi W_{\mu\nu}^{I} $ \\  
$O_{uu}$ & ($\bar{u}\gamma_{\mu}u)( \bar{u}\gamma^{\mu} u)$ & $O_{tG}$  & $(\bar{q}\sigma^{\mu \nu} \lambda^A t) \tilde\phi G^A_{\mu\nu}$ \\  
$O^{8}_{qu}$ & ($\bar{q}\gamma_{\mu}T^Aq)( \bar{u}\gamma^{\mu} T^Au)$ & $O_{G}$   & $ f_{ABC} \, G_{\mu}^{A \nu}G_{\nu}^{B \lambda} G_{\lambda}^{C \mu}$ \\  
$O^{8}_{qd}$ & ($\bar{q}\gamma_{\mu}T^Aq)( \bar{d}\gamma^{\mu} T^Ad)$  & $O_{\tilde G}$  & $ f_{ABC} \, \tilde G_{\mu}^{A \nu}G_{\nu}^{B \lambda} G_{\lambda}^{C \mu}$ \\  
$O^{8}_{ud}$ & ($\bar{u}\gamma_{\mu}T^Au)( \bar{d}\gamma^{\mu} T^Ad)$  & $O_{\phi G}$  & $(\phi^\dagger \phi)G_{\mu\nu}^{A}G^{A \mu\nu} $ \\ 
 & & $O_{\phi \tilde G}$    & $(\phi^\dagger \phi) \tilde G_{\mu\nu}^{A}G^{A \mu\nu} $  \\
\bottomrule
\end{tabular}
\caption{\label{table:opset} All dimension-six operators relevant to top
  quark production, in the notation of
  Ref.~\cite{Grzadowski:2010es}. Details of each are included in the
  text. $q$ denotes the left-handed quark doublet, $u$ and $d$ denote the up-type and down-type right-handed singlets. We do not include explicit flavor indices here, the relevant flavor indices are included in the text. 13 operators are shown,
  but $O_{tW}$ and $O_{tG}$ have both real and imaginary parts which should be
  considered as independent operators; the latter produce $\mathcal{CP}$-violating
  effects.}
\end{table}


\begin{table*}[!t]
\begin{center}
\parbox{0.55\textwidth}{
\begin{tabular}{lllc}
\toprule
Dataset & $\sqrt{s}$ (\tev) & {Measurements} & {Ref.} \\
\midrule
\multicolumn{4}{l}{\textit{Top pair production}} \\
ATLAS  & 7 + 8 & Total inclusive $\sigma$  & \cite{Aad:2014kva} \\
       & 7 + 8 & Differential $p_T (t),M_{t\bar{t}},|y(t\bar{t})|$  & \cite{Aad:2014zka} \\
CMS    & 7      & Differential $p_T (t),M_{t\bar{t}},y(t),|y(t\bar{t})|$  & \cite{Chatrchyan:2012saa} \\
CDF    & 1.96   & Differential $M_{t\bar{t}}$  & \cite{Aaltonen:2009iz} \\
D${\slashed 0}$      & 1.96  & Differential $M_{t\bar{t}},p_T(t),|y(t)|$ & \cite{Abazov:2014vga} \\
\addlinespace
\multicolumn{4}{l}{\textit{Single top production}} \\
ATLAS $t$-channel & 7  & Total inclusive $\sigma$  & \multirow{2}{*}{\cite{Aad:2014fwa}}\\
                  & 7  & Differential $p_T (t),|y(t)|$ & \\
CMS $t$-channel & 7  & Total inclusive $\sigma$  & \cite{Chatrchyan:2012ep} \\
                & 8  & Total inclusive $\sigma$  & \cite{Khachatryan:2014iya} \\
CDF  $s$-channel & 1.96  & Total inclusive $\sigma$  & \cite{Aaltonen:2014qja} \\
D${\slashed 0}$  $s+t$-channel & 1.96  & Total inclusive $\sigma$  & \cite{Abazov:2013qka} \\
\bottomrule
\end{tabular}}
\hspace{0.2cm}
\parbox{0.35\textwidth}{\vspace{4.1cm}
  \caption{\label{table:datasets} Datasets used in the fit, including total cross-sections ($\sigma$); transverse momenta of single tops ($p_T(t)$) and top pairs ($p_T(t\bar{t})$); rapidities of single tops ($y(t)$) and top pairs ($y(t\bar{t})$); and the invariant mass of top pairs ($M_{t\bar{t}}$).}}
\end{center}
\end{table*}

Given the plethora of concrete scenarios and the absence of any
telling signals of new physics in the current data, parametrising BSM
effects in an effective field theory expansion~\cite{Weinberg:1979} is
well-motivated. In this approach, all possible interactions are
captured in an effective Lagrangian~$\mathcal{L}_{\text{eff}}$:
\begin{equation*}
  \mathcal{L}_{\text{eff}}= \mathcal{L}_{\text{SM}}+\frac{1}{\Lambda}\mathcal{L}_{1}+\frac{1}{\Lambda^{2}}\mathcal{L}_{2}+ ...  \quad .
\end{equation*}
The higher-dimensional Lagrangian terms $\mathcal{L}_i$ are suppressed
by powers of $\Lambda$ - the energy scale associated with the new
physics. In the top-down approach, we have integrated out all heavy
degrees of freedom, capturing their low energy phenomenology guided by
SM gauge and global symmetries, irrespective of their concrete UV
dynamics. Such an expansion is valid provided there is a good
separation of scales between the typical collider energy and
$\Lambda$. However, this approach is completely general: the
$\{\mathcal{L}_i\}$ are constructed from SM operators, respecting the
$SU(3) \times SU(2) \times U(1)$ gauge symmetry.

The leading contributions relevant to new physics in the top sector
enter at the dimension-six level $\mathcal{O}(1/\Lambda^2)$
\begin{equation*}
  \mathcal{L}_{\text{eff}}= \mathcal{L}_{\text{SM}}+\frac{1}{\Lambda^2}\sum_{i}C_{i}O_{i} +\mathcal{O}(\Lambda^{-4})\,,
\end{equation*}
where $C_i$ are arbitrary `Wilson coefficients' and $O_{i}$ are
dimension-six operators. These operators lead to noticeable deviations
from SM expectations in a double expansion of the matrix element in SM
and new physics couplings
\begin{equation}
  |\mathcal{M}_{\text{tot}}|^2 = |\mathcal{M}_{\text{SM}}|^2
  +2\Re\left\{ \mathcal{M}_{\text{SM}} \mathcal{M}_{D6}^\ast\right\}
  + |\mathcal{M}_{D6}|^2,
\label{eqn:m2}
\end{equation}
where strictly speaking one must neglect the third term on the
right-hand side if working to dimension-six only, as this has
dimension-eight,  . Provided $C_i/\Lambda^2$ is small, such a truncation
is typically valid and the squared dimension-six terms become
numerically irrelevant.

The complete set of 80 effective operators at dimension-six has been
known for some
time~\cite{Buchmuller:1986,Burgess:1983,Leung:1986}. Only recently was
it shown that this basis contains several redundancies, with the
minimal set comprising 59 terms~\cite{Grzadkowski:2003tf,AguilarSaavedra:2008zc,AguilarSaavedra:2008zc,Grzadowski:2010es}. Considerable
attention has been devoted to constraining these operators, for
example, in the context of Higgs and precision electroweak
physics~\cite{Corbett:2015ksa,Ellis:2014jta,Contino:2013kra,Pomarol:2013zra,Efrati:2015eaa}.
In addition, strong bounds have also been placed on new top
interactions from precision constraints at LEP~\cite{Mebane:2013zga}
and direct searches for top quark physics at the
LHC~\cite{Hioki:2009hm,Hioki:2010zu,Degrande:2010kt,Englert:2012by,Hioki:2013hva,Aguilar-Saavedra:2014iga}.

While Higgs physics has received a lot of attention from an EFT perspective, the
top quark sector has not seen similar scrutiny, although top data from the
combination of the Tevatron and the LHC Run~I is far more abundant. In the last
few years, top quark physics has entered something of a precision era: the top
has been measured in several production and decay channels, and dedicated
searches in complicated final states such as $t\bar{t}H$ are
underway~\cite{Khachatryan:2015ila,Aad:2015gra}.

It is our aim to close this gap. The \textsc{TopFitter} approach
constrains new physics in the top sector using both differential and
inclusive observables, by means of a computational tool which is fully
flexible with respect to the number of input measurements and scales
well to the relevant number of EFT operators. In the present work we
limit ourselves to a nine-dimensional fit based on direct top
measurements performed at the Tevatron and the LHC, keeping track of
all EFT operator-correlations, and reserve a more complete
investigation for the near future~\cite{toapp}.

\section{Relevant Operators}
Throughout the analysis, and for ease of comparison with precision
electroweak studies, the operator set presented in
Ref.~\cite{Grzadowski:2010es} is used (see also the basis of
Refs.~\cite{ZhangWillenbrock:2011,Greiner:2011}). Assuming minimal
flavor violation, and in the leading-order\footnote{By leading-order we mean $\mathcal{O}(\Lambda^{-2})$, but for some new physics effects, such as top flavour-changing neutral currents, the first non-zero  contributions enter at $\mathcal{O}(\Lambda^{-4})$ see e.g. \cite{AguilarSaavedra:2010sq} for details.} approximation of equation 2, of these 59 operators only 15 --- shown in
Table~\ref{table:opset} --- are relevant for top production.
Fitting a 15-dimensional function is a considerable challenge; a brute
force likelihood scan at $N$ points per dimension would require
$N^{15}$ evaluations, which is prohibitive even for modest,
low-resolution values of $N$. This na\"ive dimensionality can be
reduced, however, by noting some features of the operator set.

Firstly, we note that the two operators containing the dual
field-strength tensor $\tilde G_{\mu \nu}=\epsilon_{\mu \nu \rho
  \sigma}G^{\rho \sigma}$, along with the imaginary parts of $O_{tG}$
and $O_{tW}$, are $\mathcal{CP}$-odd and can be discriminated from
${\mathcal{CP}}$-even effects in studies of spin-correlations,
polarisation effects and genuinely ${\mathcal{CP}}$-sensitive
observables \cite{Han:2009ra} (for recent analyses focusing on the
$tW\!b$ vertex, for instance, see Refs.~\cite{Chen:2005vr,AguilarSaavedra:2008gt,AguilarSaavedra:2010nx,AguilarSaavedra:2011ct,Bernardo:2014vha}). Currently there is
no evidence for ${\mathcal{CP}}$-violation in the top sector beyond
the minimal flavor violation assumption. We will address these
operators in forthcoming work but neglect them in the following; the
dimensionality of our fit is reduced by four.

Secondly, we consider top-pair production. Here the four-fermion
operators, which are numerous when all flavour combinations are
considered, only contribute to top-pair production through the
partonic subprocesses $u\bar{u},d\bar{d}\to t\bar{t}$, which reduces
the myriad of possible operators to four unique, flavour-specific
linear combinations~\cite{AguilarSaavedra:2010,ZhangWillenbrock:2011}:
\begin{equation*}
\begin{split}
C^1_u =&~3\, (2\,C^{(1)\,{1331}}_{qq}+C^{1331}_{uu}) \\
              &-(C^{(1)\,{1133}}_{qq}+C^{(3)\,{1133}}_{qq}+C^{1133}_{uu}) \\
C^2_u = &-(C^{(8)\,{1133}}_{qu} + C^{(8)\,{3311}}_{qu}) \\
 C^1_d =&~ 3\,(C^{(3)\,{1331}}_{qq} - C^{(1)\,{1331}}_{qq}) \\
              &+(C^{(3)\,{1133}}_{qq} - C^{(1)\,{1133}}_{qq})+6\,C^{(8)\,{3311}}_{ud} \\
C^2_d = &-(C^{(8)\,{1133}}_{qu} + C^{(8)\,{3311}}_{qd}),
\end{split}
\end{equation*}
 where explicit flavor indices $(\bar{q}^iq^j)(\bar{q}^jq^k)$ have now
been included. The non-4-fermion operators $O_{G}$, $O_{tG}$, and
$O_{\phi G}$ also contribute to top pair production, giving a total of
7 relevant operators. In the $gg\to t\bar{t}$ channel, $O_G$ rescales
the triple gluon vertex while $O_{tG}$ modifies the top--gluon
coupling; $O_{\phi G}$ only contributes through $gg\to h\to t\bar{t}$,
which is heavily suppressed in the Standard Model although it can be
probed in $ttH$ production.

Three $\mathcal{CP}$-even operators\footnote{The contribution of the operator $O^1_{\phi q} = (\phi^\dagger D_\mu \phi) (\bar{t} \gamma^\mu b)$ is heavily suppressed, as its interference with the SM amplitude is proportional to $m_b$ (see e.g.~\cite{Cao:2007ea}).} contribute to single top
production: $O_{tW}$ modifies the $tW\!b$ vertex, as does $O^3_{\phi
  q}$, while the operator $O^{(3)\,{1331}}_{qq}$ creates a new
four-quark topology which interferes with the SM piece.

There is hence a clean factorisation into 7+3 $\mathcal{CP}$-even
operators associated with top quark production at hadron colliders. In
this study we reduce this further to a 6+3 configuration by eliding
the highly suppressed contributions of $O_{\phi G}$ to top-pair
production.

\section{Datasets}
The aim of this paper is to present a preliminary study demonstrating
the feasibility of performing a full global fit of top quark effective
theory to data. We thus include top quark pair and single top
production processes at parton level only, which observables and
datasets~\cite{Aad:2014zka,Chatrchyan:2012saa,Aaltonen:2009iz,Abazov:2014vga,Aad:2014fwa,Chatrchyan:2012ep,Khachatryan:2014iya,Aaltonen:2014qja,Abazov:2013qka}
are collected in Table \ref{table:datasets}. 

It should be noted that a fully differential fit along these lines consists of a multitude of exclusive measurements. Treating each bin as an independent\footnote{Where published by the experiments, we have included bin-to-bin correlations. These have negligible effect on our conclusions. }  measurement, we have 103 bins for top pair production and 23 from single top. This highlights the necessity of a fast analysis framework, as introduced in the present work. 

Given that we will model
higher-order corrections as described in the following section, we do
not include $Wt$ production, which interferes with top pair production
at next-to leading order, such that it is not possible to reproduce
existing experimental analyses using a fixed order parton level
calculation~\cite{Zhu:2001hw,Kauer:2001sp,Campbell:2005bb,Frixione:2008yi,White:2009yt}.

\section{Details of Analysis}
We begin by including the operators listed above (together with
consequent SM parameter redefinitions) in a
\textsc{FeynRules}~\cite{Christensen:2008py} model file, which is then
interfaced via UFO~\cite{Degrande:2011ua} to
\textsc{MadGraph/MadEvent}~\cite{Alwall:2011uj,Degrande:2011ua} in
order to obtain parton-level theory predictions. Samples were
generated for all the relevant processes: top pair production: $pp \to
t\bar{t}$, single top production: $pp \to t\bar{b}$ ($s$-channel), and
$pp \to tq$ ($t$-channel).

In order to model next-to leading order QCD corrections, SM-only
samples at next-to-leading order are generated with
\textsc{Mcfm}~\cite{Campbell:2010ff}. These are used to construct
differential (bin-by-bin) and global $K$-factors, as in
e.g. Ref.~\cite{Englert:2014oea}. Theoretical uncertainties for these
samples are estimated in the usual way, by independently varying the
scales $\mu_{\text{central}}/2 < \mu_\mathrm{R,F} <
2\mu_{\text{central}}$, where $\mu_{\text{central}}$ is taken to be
$m_t$. Parton distribution function (PDF) uncertainties are estimated
by generating events using the next-to-leading order
NNPDF23~\cite{Ball:2010de}, MSTW2008~\cite{Martin:2009iq}, and
CT10~\cite{Nadolsky:2008zw} PDF sets, according to the
PDF4LHC~\cite{Botje:2011sn} prescription. We take the central value as
our estimate and the width of the envelope (including scale
variations) as the total theoretical uncertainty. In the case of top
pair total inclusive cross-sections, we use global $K$-factors from
next-to-next-to leading order QCD with soft gluons resummed to
next-to-next-to-leading logarithmic
accuracy~\cite{Beneke:2011mq,Cacciari:2011hy,Czakon:2011xx,Baernreuther:2012ws,Czakon:2013goa}.

A strength of our fitting procedure is the use of novel techniques
developed in the context of Monte Carlo event generator tuning, as
implemented in the \textsc{Professor}~\cite{Buckley:2009bj}
framework. The procedure is as follows:

\begin{figure}[!t]
\begin{center}
\includegraphics[width=0.45\textwidth]{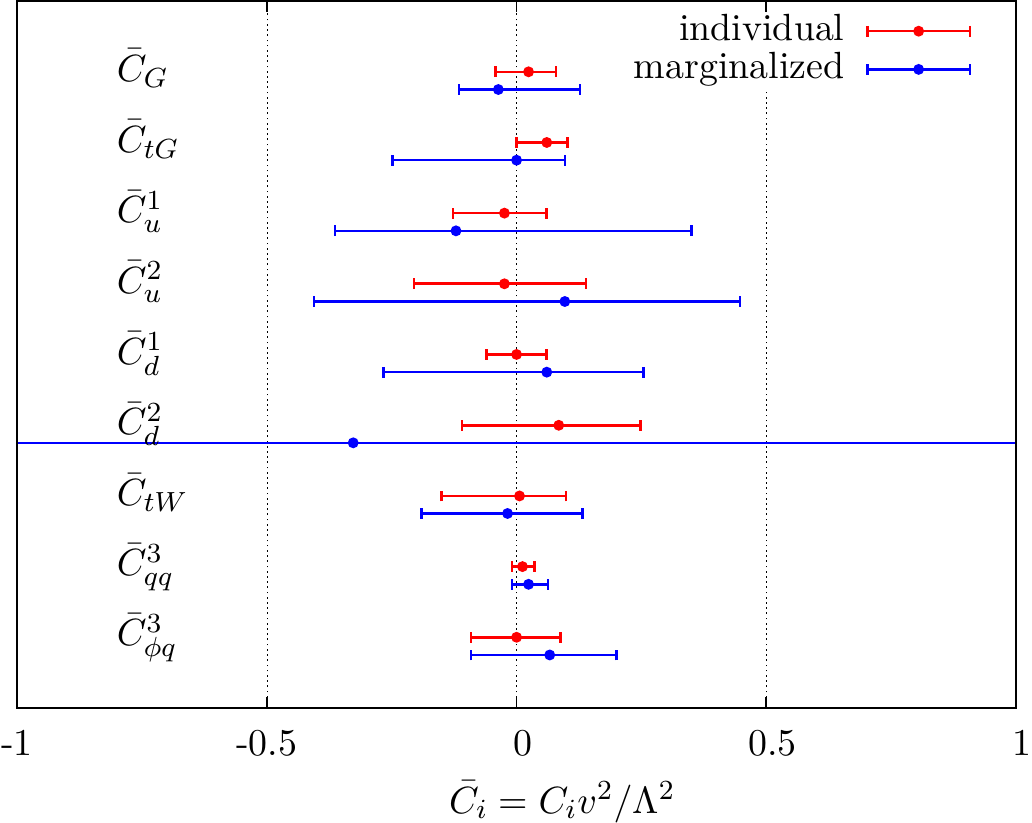}
\caption{95\% confidence intervals for operators contributing to
  top-pair and single top production, individually (with all other
  operators set to zero) and marginalised (with all other operators
  allowed to float to best-fit values). Note that the marginalised bound on $\bar{C^2_d}$ fall outside the region where the dimension-six approximation is valid, so this operator is unconstrained.}
\label{fig:ops}
\end{center}
\end{figure}

\begin{figure*}[!t]
\begin{center}
  \subfigure[~individual constraints.\label{fig:contour-indiv}]{\includegraphics[width=0.42\textwidth]{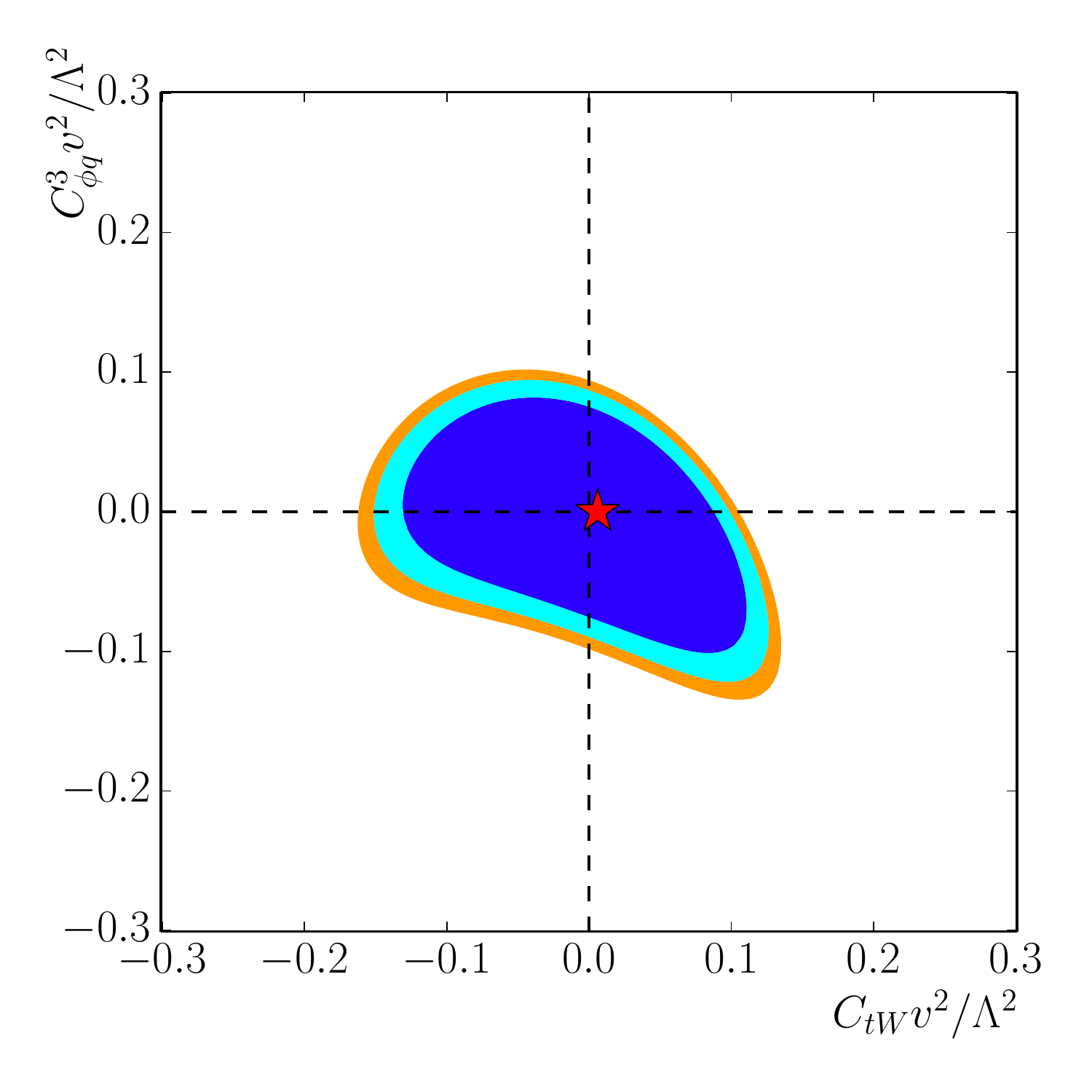}}
  \hspace{1cm}
  \subfigure[~marginalised constraints.\label{fig:contour-marg}]{\includegraphics[width=0.42\textwidth]{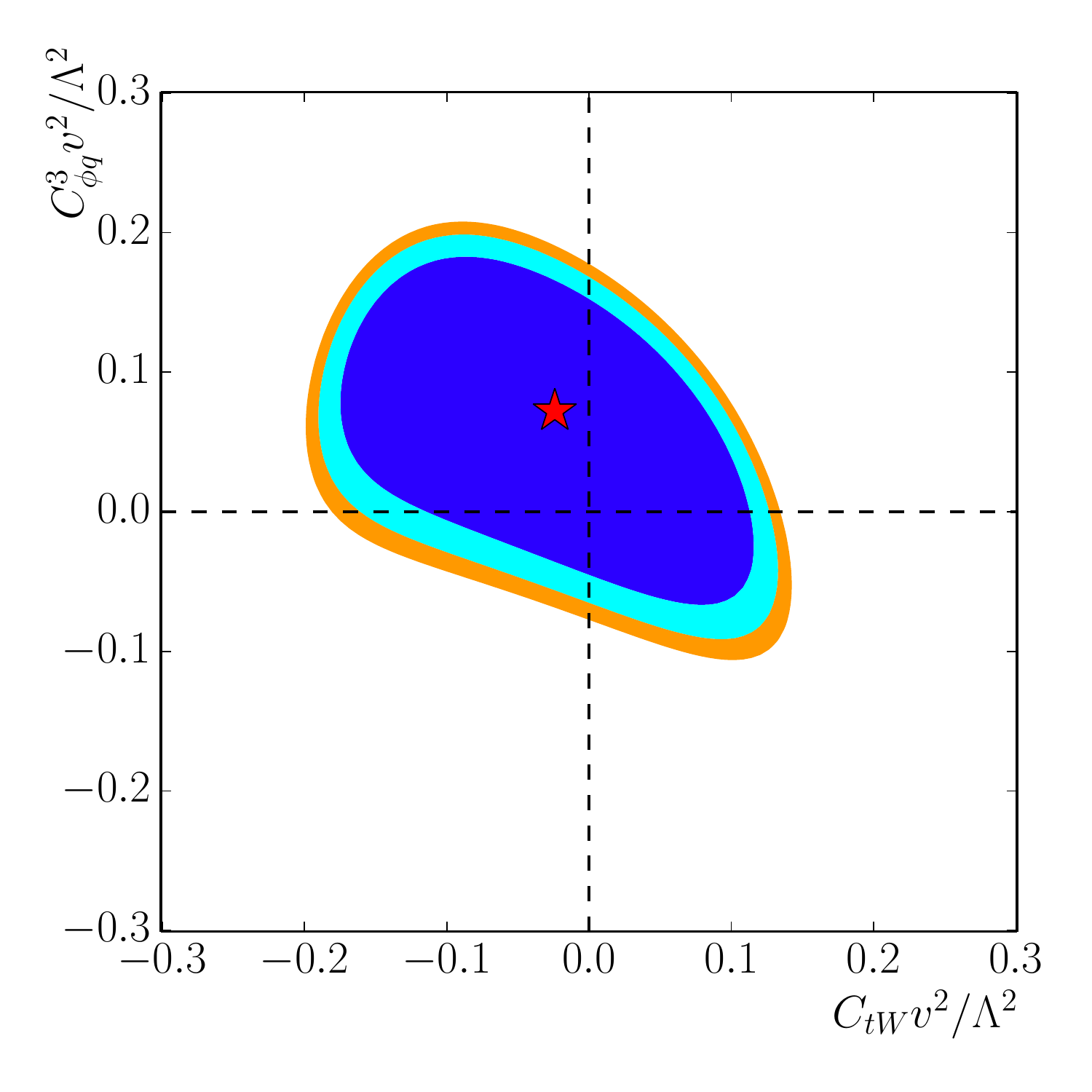}}
  \caption{68\% (blue), 95\% (turquoise) and 99\% (orange) confidence
    intervals for $C_{tW}$ and $C^3_{\phi q}$ in a global fit, with all
    remaining coefficients set to zero (a) and marginalised over
    (b). The star marks the best fit point, indicating a currently
    good agreement with the Standard Model.}
\label{fig:contour}
\end{center}
\end{figure*}

\begin{itemize}
\item A set of points in the $N$-dimensional parameter space $\{C_i\}$
  is sampled logarithmically. Other samplings are possible - we choose
  logarithmic sampling to avoid oversampling of regions where
  coefficients are large, such that dimension-eight terms become
  important.
\item At each sampled parameter space point, all theory observables are
  calculated, with uncertainties, as described above. One then constructs a
  polynomial \textit{parameterising function} $f_b(\{C_i\})$ for each observable
  bin $b$, which fits the sampled points with least-squares-optimal
  precision. This function can be used to efficiently generate theory
  predictions for arbitrary parameter space points within the fitted range. We
  choose a third-degree polynomial for this function. This has been shown to
  work well in Monte Carlo tuning~\cite{Buckley:2009bj}, and should in fact be
  better suited to the present case: in the absence of uncertainties, each
  observable is a second-order polynomial in the $\{C_i\}$,
  cf. eq.~\eqref{eqn:m2}, and the extra polynomial order provides some tolerance
  to beyond-fixed-order effects.
\item Finally, we construct a $\chi^2$ function between the bin
  parameterisations $\{f_b(\{C_i\})\}$ and the data, according to
\begin{equation*}
  \chi^2(\{C_i\}) = \sum_{\substack{\mathcal{O}} } \sum_{\substack{b} } \frac{(f_b(\{C_i\}) - E_b)^2}{\sigma_b^2} ,
\end{equation*}
i.e. we sum over all observables $\mathcal{O}$, and all bins in that
observable, $b$. $E_b$ is the experimental reference value at bin $b$
and $\sigma_b$ is the total uncertainty for bin $b$, which we for now
assume as an uncorrelated combination of theoretical modelling and
experimental measurement uncertainties, $\sigma_b =
\sqrt{\sigma_{\text{theory}}^2+\sigma_{\text{exp}}^2}$. The $\chi^2$
is then used to place constraints on the operator Wilson coefficients,
as follows.
\end{itemize}
Constraints are obtained in two ways, for ease of comparison with
existing literature. Firstly, single operator coefficients are allowed
to vary, with all others set to zero (the SM value). The $\chi^2$ is
then minimised using {\sc PyMinuit}~\cite{James:1975dr}, and used to
set confidence limits on the operator value. A second approach is to
marginalise over the remaining operators, namely to construct the
confidence limit for a given operator coefficient whilst allowing all
other coefficients to vary. Both cases are shown in
Figure~\ref{fig:ops}, where the dimension-six contributions are
normalised to the Standard Model piece via
$\bar{C_i}=C_iv^2/\Lambda^2$.  All results are consistent with the SM
within 95\% limits. 

As with all effective operator constraints, these must be interpreted as valid only in the region where $\mathcal{O}(\Lambda^{-4})$ terms are not large. Clearly $\bar{C^2_d}$ is outside this region. In top pair production, for instance, the contribution from dimension-6 operators relative to the SM piece is typically $\mathcal{O}(g_s^2 C_iv^2/\Lambda^2)$ which must be $<$ 1 in the linear approximation , i.e $\bar{C_i} \lesssim 1.5$. All other operators respect this bound. It should be noted that some of these operators, namely those containing field strength tensors, can only be generated at loop level in the ultraviolet completion, which widens this region of validity since $\Lambda^2$ will be accompanied by a loop factor of $16\pi^2$. This argument is invalid, however, if the underlying completion is strongly coupled. It is possible to include such information in our fitting approach, but in the interests of full generality no such model-specific assumptions are made here.

One sees from Figure~\ref{fig:ops} that the weakest constraints are on the
coefficients (of four-fermion operators) $\bar{C}_u^i$ and
$\bar{C}_d^i$. These are constrained by the processes
$u\bar{u}\rightarrow t\bar{t}$ and $d\bar{d}\rightarrow t\bar{t}$
respectively, which are suppressed relative to the corresponding gluon
initiated processes, mostly due to the relative partonic luminosities.

One may also examine the correlation of constraints between pairs of
operators. An example is Figure~\ref{fig:contour-indiv}, which shows
confidence limits in the $(C_{tW},C^3_{\phi q})$ plane, with all other
operator coefficients set to zero. One may also marginalise over all
remaining operators, as shown in Figure~\ref{fig:contour-marg}. In
both cases, we currently find excellent agreement with the SM. More
detailed results will be presented in a forthcoming
paper~\cite{toapp}.

\section{Summary, Conclusions and Outlook}
\label{sec:conclude}
Following the discovery of the Higgs boson, the search for physics
beyond the Standard Model will remain the primary goal of the LHC
experiment for the foreseeable future. The top quark sector is a
particularly well-motivated window through which to look for the
imprint of non-resonant new physics. Modelling such effects using
effective field theory (higher dimensional operators) is well
justified given the absence of new resonant physics from the LHC
Run~I. The abundance of top quark production at the LHC enables a
multi-faceted analysis of top quark phenomenology and allows us to
confront higher dimensional top sector operators with differential
measurements at high statistics.

In this paper, we have characterised new physics corrections using the
well-established framework of effective field theory. We have
presented results from a new computational framework to fit all
possible dimension-six operator coefficients to a comprehensive set of
relevant data. This is possible through our use of fast-fitting
algorithms, which have been developed (and well-tested) in the context
of Monte Carlo event generator tuning. Here we expect these techniques
to work even better, given the explicit polynomial dependence of
theory observables on operator coefficients.

Our method involves constructing a parameterising function to
effectively parametrise the theory output of Monte Carlo generators
(here at parton level only). Once this has been constructed, it is
quick to perform a global fit containing all possible operators, and
to amend this fit as and when new data appear. Furthermore, there is
no significant speed decrease in our fitting procedure upon improving
the theory prediction (e.g. to include parton shower or detector
corrections), as such improvements only affect the parameterising
function, which has to be calculated only once.

The results of our fit currently show good agreement with the Standard
Model, which is unsurprising given the absence of new physics
currently reported in other studies. Our results, however, provide a
proof of principle study that efficient global fits of top quark
effective theory are possible. It is straightforward to generalise our
fit to include more experimental observables (beyond parton level,
including top quark decays), to improve the theory description with
higher order corrections, and to include new data sets including those
from the recently commenced LHC Run~II. Work in these directions is
ongoing.

\bigskip
\noindent
\textit{Acknowledgements} --- AB is supported by a Royal Society
University Research Fellowship. CE is supported by the IPPP
Associateship programme. DJM, LM, MR and CDW are supported by the UK
Science and Technology Facilities Council (STFC) under grant
ST/L000446/1. JF is supported under STFC grant ST/K001205/1.

\bibliography{full_ref}


\end{document}